\documentclass[fdp,a4paper,fleqn]{w-art}
\usepackage{times,cite,w-thm}    
\usepackage{graphicx}
\usepackage{amssymb}
\usepackage{hyperref}     

\begin{document}
%%    The information for the title page will be placed between
%%    \begin{document} and \maketitle. The order of most entries
%%    is determined by the class file and can not be changed by
%%    rearranging them. The maketitle command follows after the
%%    abstract.
%%
%%    Most of the following commands will be completed by the publisher.
%%
%%    The copyrightyear is defined in the .clo file as the first argument
%%    of the copyrightinfo command. If the copyrightyear differs from that
%%    value it might be adjusted by the following definition:
%%
%% \renewcommand{\copyrightyear}{2003}% uncomment to change the copyrightyear.
%%

%%%%%   macros for  LHCb 
%%
%% Luminosity
\newcommand{\lum}{\mathcal{L}}
\newcommand{\lumin}[2]{{#1}\;10^{#2}\;\mathrm{cm^{-2} s^{-1}}}
\newcommand{\lolumi}{\mbox{$2 \times 10^{32}$~cm$^{-2}$ s$^{-1}$~}}
\newcommand{\hilumi}{\mbox{$5 \times 10^{32}$~cm$^{-2}$ s$^{-1}$~}}
%% Units 
\newcommand{\cms}{\mbox{cm$^{-2}$ s$^{-1}$~}}
\newcommand{\mq}{\ensuremath{\; \mathrm{m^2}}}
\newcommand{\cmq}{\ensuremath{\; \mathrm{cm^2}}}
\newcommand{\ccm}{\ensuremath{\; \mathrm{cm^3}}}
\newcommand{\mmq}{\ensuremath{\; \mathrm{mm^2}}}
\newcommand{\us}{\; \mu\:\mathrm{s}}
\newcommand{\gra}{{\mathrm{^o}}} 
\newcommand{\bb}{$b\overline{b}$}
\newcommand{\cc}{$c\overline{c}$}
\newcommand{\invcmq}{\ensuremath{\mathrm{/{cm^2}}}}
\newcommand{\mHzpercmq}{\ensuremath{\mathrm{MHz/cm^2}}}
\newcommand{\kHzpercmq}{\ensuremath{\mathrm{kHz/cm^2}}}
\newcommand{\Hzpercmq}{\ensuremath{\mathrm{Hz/cm^2}}}
\newcommand{\Cpercm}{\ensuremath{\mathrm{C/cm}}}
\newcommand{\Cpercmq}{\ensuremath{\mathrm{C/cm^2}}}
\newcommand{\ItwoC}{\ensuremath{\mathrm{I^2C}}}
\newcommand{\x}{\ensuremath{\times}}
\newcommand{\microns}{\ensuremath{\mathrm{\mu m}}}
\newcommand{\Jpsi}{\ensuremath{J\mskip -2.5mu/\mskip -1mu\psi\mskip 1mu}}
\newcommand{\fromb}{\ensuremath{\Jpsi~\mathrm{from}~b}}
\newcommand{\microb}{\ensuremath{\,\mu\mathrm{b}}}
\newcommand{\gevc}{\ensuremath{\,\mathrm{GeV}\mskip -2mu/\mskip -1mu c}}
\newcommand{\mevc}{\ensuremath{\,\mathrm{MeV}\mskip -2mu/\mskip -1mu c}}
\newcommand{\mevcc}{\ensuremath{\,\mathrm{GeV}\mskip -2mu/\mskip -1mu c^2}}
\newcommand{\gevcc}{\ensuremath{\,\mathrm{GeV}\mskip -2mu/\mskip -1mu c^2}}
\def\PT{p_{\mathrm{T}}\xspace}
\def\ET{E_{\mathrm{T}}\xspace}
%%%
\def\Y#1S{{\Upsilon\mathrm{(#1S)}}}
\def\pipi{\pi^+\pi^-}
\def\ubar{\bar u}
\def\dbar{\bar d}
\def\sbar{\bar s}
\def\cbar{\bar c}
\def\bbar{\bar b}
\def\tbar{\bar t}
\def\Ks{K^0_S}
\def\Kl{K^0_L}
\def\BS{$B_s$\xspace}
\def\BC{$B_c$\xspace}
\def\BBar{$B^0_s$-${\kern 0.18em\overline{\kern -0.18em B^0_s}}$\xspace}
%%%%%%
\def\BJMUK{$B^+ \rightarrow \Jpsi(\mu^+ \mu^-) K^+$}
\def\BSJMUPHIK{$B^0_s \rightarrow \Jpsi(\mu^+ \mu^-)\phi(K^+K^-)$\xspace}
\def\BKPI{$B^0 \rightarrow K^+ \pi^-$\xspace}
\def\BJMUK{$B^0_d \rightarrow \Jpsi(\mu^+ \mu^-) K^0_S$}
\def\BJMUKVIS{$B^0_d \rightarrow \Jpsi(\mu^+ \mu^-) K^0_S(\pi^+ \pi^-)$}
\def\BKMUMU{$B_d \rightarrow K^{*0}\mu^+ \mu^-$}
\def\BJMUKBOLD{$\bfmath{B^0_d \rightarrow \Jpsi(\mu^+ \mu^-) K^0_S}$}
\def\BJPSIPHI{$B^0_s \rightarrow \Jpsi \phi$\xspace}
\def\BJPHI{$\rm B^0_s \rightarrow \Jpsi \phi$\xspace}
\def\BSMUMU{$B^0_s \rightarrow \mu^+ \mu^-$\xspace}
\def\BMUMU{$B \rightarrow \mu^+ \mu^-$\xspace}
\def\B0MUMU{$B^0 \rightarrow \mu^+ \mu^-$\xspace}
\def\BSMUMUBOLD{$\bfmath{B^0_s \rightarrow \mu^+ \mu^-}\xspace$}
\def\BJPSIKS{$B^0_d \rightarrow \Jpsi Ks $}
\def\BMUX{$\rm b \rightarrow \mu\ X$}
\def\pkm{$\rm \pi,K \rightarrow \mu\ \nu$}

\DOIsuffix{theDOIsuffix}
%%
%% issueinfo for header and copyright line
\Volume{55}
\Issue{1}
\Month{04}
\Year{2011}
%%
%%    First and last pagenumber of the article. If the option
%%    'autolastpage' is set (default) the second argument may be left empty.
\pagespan{3}{}
%%
%%    Dates will be filled in by the publisher. The 'reviseddate' and
%%    'dateposted' (Published online) entry may be left empty.
\Receiveddate{XX April 2011}
\Reviseddate{XX April 2011}
\Accepteddate{XXX 2011}
\Dateposted{XXX 2011}

%% \pretitle{Editor's Choice}

%% We have a short and a long form for the title. The short form
%% (optional argument) goes into the running head.

\title[LHCb: detector performance and first physics results]{LHCb: detector performance and first physics results}

%% Please do not enter footnotes or \inst{}-notes into the optional
%% argument of the author command. The optional argument will go into
%% the header.  If there is only one address the marker \inst{x} may be
%% omitted.

%% Information for the first author.
\author[M. Pepe Altarelli]{Monica Pepe Altarelli\inst{1}%
  \footnote{Corresponding author\quad E-mail:~\textsf{Monica.Pepe.Altarelli@cern.ch}, 
            Phone: +00\,41\,22\,7674473}} 
%%           Fax: +00\,999\,999\,999}}
\address[\inst{1}]{CERN, PH Department,  CH-1211 Geneve 23, Switzerland}
%%
%%    Information for the second author
%%\author[S. Author]{Second Author\inst{1,2,}\footnote{Second author footnote.}}
%%\address[\inst{2}]{Second address}
%%
%%    Information for the third author
%%\author[T. Author]{Third Author\inst{2,}\footnote{Third author footnote.}}
%%
%%    \dedicatory{This is a dedicatory.}
\begin{abstract}
LHCb is a dedicated detector for  $b$ and $c$ physics at the LHC. I will present a concise review of the detector design and performance together with a selection of early physics results and prospects. The integrated luminosity of $37~\rm{pb}^{-1}$ collected in 2010 has already allowed LHCb to perform a number of a very significant measurements, while the data expected for 2011 have the potential of revealing New Physics effects in the  \BS sector.
\end{abstract}
%% maketitle must follow the abstract.
\maketitle                   % Produces the title.

%% If there is not enough space inside the running head
%% for all authors including the title you may provide
%% the leftmark in one of the following three forms:

%% \renewcommand{\leftmark}
%% {F. Author: A short title}

%% \renewcommand{\leftmark}
%% {F. Author and S. Author: A short title}

%% \renewcommand{\leftmark}
%% {F. Author et al.: A short title}

%% \tableofcontents  % Produces the table of contents.
\section{Introduction}
LHCb is a dedicated $b$ and $c$-physics precision experiment at the LHC searching for New Physics (NP) beyond the Standard Model (SM) through the study of very rare decays of beauty and charm-flavoured hadrons and precision measurements of CP-violating observables.  In the last decade, experiments at $B$ factories have confirmed that the mechanism proposed by Kobayashi and Maskawa is the major source of CP violation  observed so far. The SM description of flavour-changing processes has been confirmed in the $b\to d$ transition at the level of 10-20\% accuracy. However, NP effects can still be large in $b\to s$ transitions, modifying the \BS mixing phase  $\phi_s$, measured from \BJPSIPHI decays, or in channels dominated by other loop diagrams, such as the very rare decay \BSMUMU. Therefore, the challenge of current and future $b$-physics experiments is to widen the range of measured decays, reaching channels that are strongly suppressed in the SM and, more generally, to improve the precision of the measurements to achieve the necessary sensitivity to NP effects in loops. LHCb will extend the $b$-physics results from the $B$ factories by studying decays of heavier $b$ hadrons, such as \BS or \BC, which are copiously produced at the LHC. It will complement the direct search of NP at the LHC by providing important information on the NP flavour structure through a dedicated detector, optimized for this kind of physics.

\section{$b$ physics at the LHC: environment, background, general trigger issues }
The LHC is the world's most intense source of $b$ hadrons. The  \bb\ cross section in proton-proton collisions at  $\sqrt{s}=7$~TeV is measured to be $\sim 300~ \mu$b, implying that more than $10^{11}$ \bb\ pairs are produced in a standard ($10^7$s) year of running at the LHCb operational luminosity of  \lolumi.  As in the case of the Tevatron, a complete spectrum of $b$ hadrons is available, including \BS, \BC mesons and $b$ baryons, such as $\Lambda_{\rm b}$. However, less than 1\% of all inelastic events contain $b$ quarks, hence triggering is a critical issue.

At the nominal LHC design luminosity of $10^{34}$\cms, multiple $\it pp$ collisions within the same bunch crossing (so-called pile-up) would significantly complicate the $b$ decay-vertex reconstruction and flavour tagging. For this reason the luminosity at LHCb is locally controlled by displacing the beams in the vertical direction to yield  $\lum = 2-5\times10^{32}\cms$, at which, for the 1296 colliding bunches expected in 2011,  the pile-up will be between 1 and 2. Furthermore, running at relatively low luminosity reduces the detector occupancy of the tracking systems and limits radiation damage effects. 

The dominant  production mechanism at the LHC is through gluon-gluon fusion in which the momenta of the incoming partons are strongly asymmetric in the $\it pp$ centre-of-mass frame. As a consequence, the \bb\ pair is boosted along the direction of the higher momentum gluon, and both $b$ hadrons are produced in the same forward (or backward) direction in the $\it pp$ centre-of-mass frame. The detector is therefore designed as a single-arm forward spectrometer covering the pseudorapidity range $1.9<\eta<4.9$ , which ensures a high geometric efficiency for detecting all the decay particles from one $b$ hadron together with the decay particles from the accompanying \bb\ hadron to be used as a flavour tag. A modification to the LHC optics, displacing the interaction point by 11.25 m from the centre, has permitted maximum use to be made of the existing cavern by freeing 19.7~m for the LHCb detector components. 

A detector design based on a forward spectrometer offers further advantages: $b$ hadrons are expected to have a hard momentum spectrum in the forward region; their average momentum is  $\sim80~\gevc$, corresponding to approximately 7~mm mean decay distance, which facilitates the separation between primary and decay vertices. This property, coupled to the excellent vertex resolution capabilities, allows proper time to be measured with a resolution of $\sim$50 fs, which is crucial for studying CP violation and oscillations with \BS mesons, because of their high oscillation frequency. Furthermore, the forward, open geometry allows the vertex detector to be positioned very close to the beams and facilitates detector installation and maintenance. In particular, the silicon detector sensors, housed, like Roman pots, in a secondary vacuum, are split in two halves that are retracted by   $\sim$30 mm from the interaction region before the LHC ring is filled, in order to allow for beam excursions during injection and ramping. They are then positioned with their sensitive area  $\sim$8~mm from the beam during collisions. 

The LHCb acceptance in the plane $(\eta,\PT)$ of the $b$ hadrons nicely complements that of ATLAS and CMS:  ATLAS and CMS cover a pseudorapidity range of  $|\eta|<2.5$ and rely on high-$\PT$ lepton triggers. LHCb relies on much lower $\PT$ triggers, which are efficient also for purely hadronic decays. Most of the ATLAS and CMS $b$-physics programme will be pursued during the first few years of operation, for luminosities of order $10^{33}$\cms. Once LHC reaches its design luminosity, $b$ physics will become exceedingly difficult for ATLAS and CMS due to the large pile-up (20 interactions per bunch crossing, on average), except for very few specific channels characterized by a simple signature, like \BSMUMU. 

\section{Detector description and performance}
\begin{figure} [htb]
\centering
\includegraphics [width=10.0 cm]{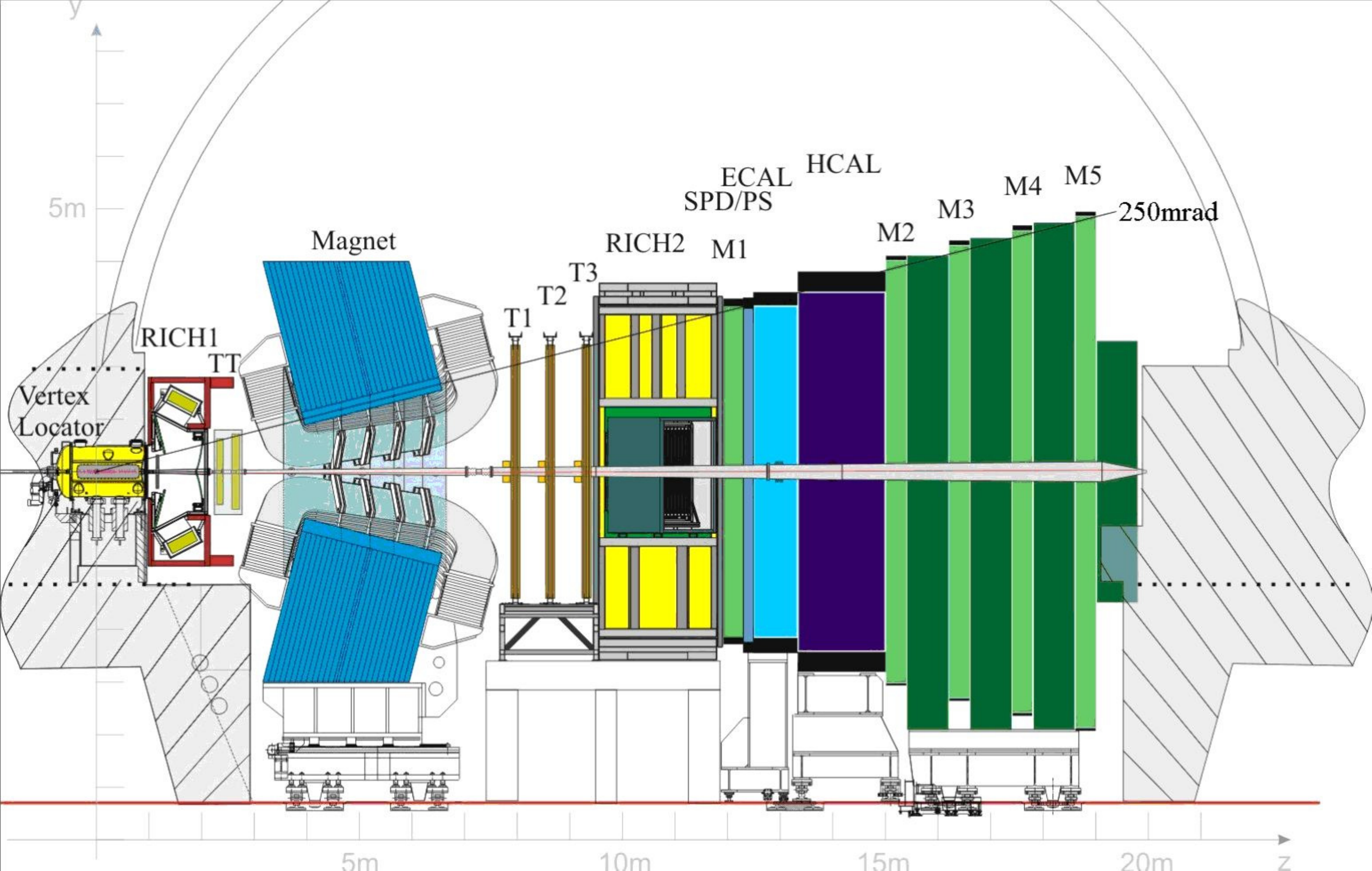}    
\caption{Side view of the LHCb detector showing the Vertex Locator (VELO), the dipole magnet, the two RICH detectors, the four tracking stations TT and T1-T3, the Scintillating Pad Detector (SPD), Preshower (PS), Electromagnetic (ECAL) and Hadronic (HCAL) calorimeters, and the five muon stations M1-M5.\label{detector}}
\end{figure}

charm 
The key features of the LHCb detector are: 
\begin{itemize}
\item A versatile trigger scheme efficient for both leptonic and hadronic final states, which is able to cope with a variety of modes with small branching fractions;
\item Excellent vertex and proper time resolution;
\item Precise particle identification, specifically for hadron ($\pi$/K) separation;
\item Precise invariant mass reconstruction to efficiently reject background due to random combinations of tracks. This implies a good momentum resolution.
\end{itemize}
A schematic layout is shown in Fig.~\ref{detector}. It consists of a vertex locator (VELO), a charged-particle tracking system with a large aperture dipole magnet, aerogel and gas Ring Imaging Cherenkov counters (RICH), electromagnetic (ECAL) and hadronic (HCAL) calorimeters and a muon system.
In the following, the most salient features of the LHCb detector are described in more detail. A much more complete description of the detector characteristics can be found in \cite{ref:reopt, ref:det}.
\subsection {Trigger}
One of the most critical elements of LHCb is the trigger system. At the chosen LHCb nominal luminosity, taking into account the LHC bunch crossing structure, the rate of events with at least two particles in the LHCb acceptance is $\sim$10~MHz (instead of the nominal 40 MHz LHC crossing rate). The rate of events containing $b$ quarks is $\sim$100~kHz, while the rate of events containing $c$ quarks is much larger. However, the rate of ÔinterestingÕ events is just a very small fraction of the total rate ($\sim$Hz), due to the combined effect of branching fractions and detector acceptance, hence the need for a highly selective and efficient trigger.

The LHCb trigger exploits the fact that $b$ hadrons are long-lived, resulting in well separated primary and secondary vertices, and have a relatively large mass, resulting in decay products with large $\PT$. It consists of two levels: Level0 (L0) and High Level Trigger (HLT). L0, implemented on custom electronics boards, is designed to reduce the input rate to 1 MHz, the rate at which all of the LHCb's subdetectors can be read out,  at a fixed latency of 4~$\mu$s. Events are then sent to a computer farm with up to $\sim$1350 multi-processor boxes where $\sim$20,000 copies of the HLT software algorithm are executed in parallel. The HLT reduces the rate from 1 MHz to 2-3~kHz.  

L0, based on calorimeter and muon chamber information, selects muons, electrons, photons or hadrons above a given $\PT$ or $\ET$ threshold, typically in the range 1 to 4 GeV.  The plan for the 2011 run, with full luminosity, is to select ~400 kHz of the bandwidth with the L0 hadron trigger, which is unique within the LHC experiments, and  $\sim$400 kHz for the muon triggers (single and double), while the rest of the bandwidth will be occupied by the electromagnetic calorimeter triggers. 

The HLT algorithms are designed to be simple, to minimize systematic uncertainties, inclusive and fast. This is realized by performing at the first stage of the HLT, called  HLT1, a partial reconstruction to select a single, good-quality track with high momentum and large impact parameter, as well as lifetime-unbiased muons and electrons. The second stage of the HLT, called HLT2, processes few enough events so that it is possible to perform a reconstruction very similar to the offline one. The average HLT execution time is O(20~ms) per event. In the 2011 run, the total trigger rate after the HLT is $\sim$3~kHz, a relatively high rate exceeding the design value of 2~kHz. The reason for this increase is the broader physics programme that the experiment is now pursuing, in particular in the area of charm decays. The experiment will collect clean samples of approximately 1 kHz each of $b$ decays to leptons, $b$ decays to hadrons and charm.

\subsection{ VELO and Tracking System}		
The LHCb tracking system consists of a warm dipole magnet, which generates a magnetic field integral of $\sim$4~Tm, four tracking stations and the VELO. The first tracking station located upstream of the magnet consists of four layers of silicon strip detectors. The remaining three stations downstream of the magnet are each constructed from four double-layers of straw tubes in the outer region, covering most ($\sim$98\%) of the tracker area, and silicon strips in the area closer to the beam pipe ($\sim$2\%). However, $\sim$20\% of the charged particles traversing the detector go through the silicon inner tracker, due to the forward-peaked multiplicity distribution. The measured impact-parameter resolution is $14~\mu$m in the highest $\PT$ bin, in good agreement with Monte-Carlo expectations.

The VELO consists of 21 stations, each made of two silicon half disks and variable pitch size, which measure the radial and azimuthal coordinates. A single hit resolution of $4~\mu$m has been achieved at optimal projected track angle for the smallest pitch size.  The VELO has the unique feature of being located at a very close distance from the beam line (0.8 cm), inside a vacuum vessel, separated from the beam vacuum by a thin aluminum foil. This allows an impressive vertex resolution to be achieved, translating, for instance, in a proper time resolution of $\sim$50~fs for the decay \BJPSIPHI, $\it{i.e.}$, a factor of seven smaller than the \BS oscillation period. Primary vertex resolutions have been measured by randomly splitting the reconstructed tracks into two subsets and by reconstructing vertices from each of the subsets. For a typical primary vertex producing 25 tracks, the resolution is  $15~\mu$m in X and Y, and $75~\mu$m in Z .
%%The resolution on the impact parameter can be parameterized as $\delta{IP}\sim14\mu\rm{m} + 35\mu\rm{m}/\PT$.
\subsection{Particle Identification}
Particle identification is provided by the two RICH detectors and the Calorimeter and Muon systems. The RICH system is one of the crucial components of the LHCb detector.  The first RICH, located upstream of the magnet, employs two radiators, $\rm C_4 F_{10}$ gas and aerogel, ensuring a good separation between kaons and pions in the momentum range from 2 to $60~\gevc$. A second RICH in front of the calorimeters, uses a  gas radiator, $\rm CF_4$, and extends the momentum coverage up to $\sim100~\gevc$. The calorimeter system comprises a pre-shower detector consisting of 2.5 radiation length lead sheet sandwiched between two scintillator plates, a 25 radiation length lead-scintillator electromagnetic calorimeter of the shashlik type and a 5.6 interaction length iron-scintillator hadron calorimeter. The muon detector consists of five muon stations equipped with multiwire proportional chambers, with the exception of the centre of the first station, which uses triple-GEM detectors. 

The identification of electrons, photons and $\pi^0$s is based on the balance of the energy deposited in the calorimeter system and the track momentum, and the matching between the barycenter position of the calorimeter cluster and the extrapolated track impact point. The average electron identification efficiency  extracted from electrons produced in photon conversions is $\gtrsim90\%$ with a misidentification rate of $\sim$3-5\% for electrons with momentum above $10~\gevc$. Muons are identified by extrapolating well reconstructed tracks with p$\ge3~\gevc$ into the muon stations and matching the tracks with the hits within the corresponding fields of interest. The efficiency in the acceptance extracted from $\Jpsi\rightarrow \mu\mu$ decays is $\sim97\%$. The $\mu$-$\pi$ and $\mu$-$\rm{K}$ misidentification rates are dominated by $\pi$ and K decays in flight and are found to be below $1\%$ for momenta above $20~\gevc$. 

\begin{figure} [htb]
\centering
\includegraphics[width=68mm,height=55mm]{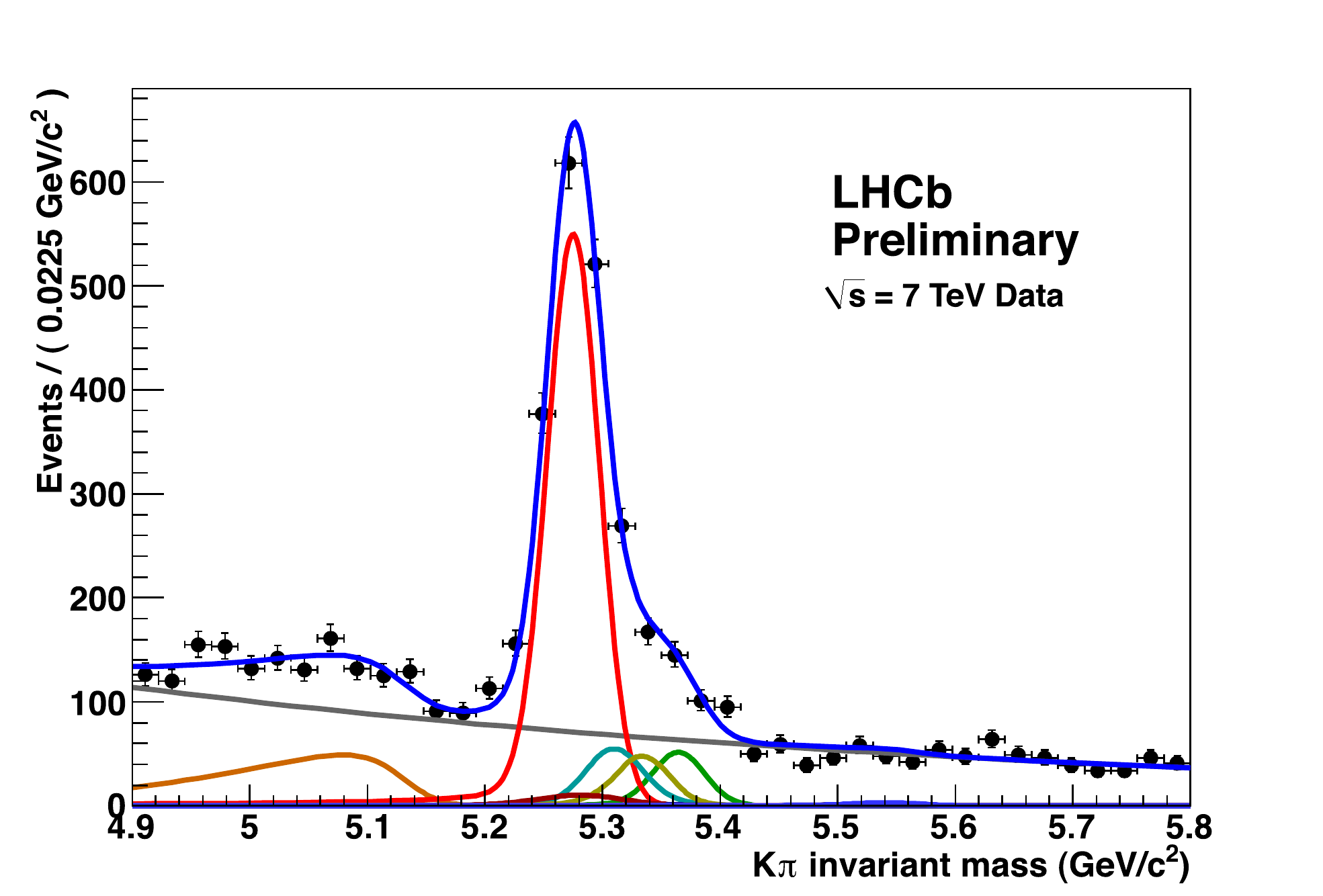}
\hfil
\includegraphics[width=68mm,height=55mm]{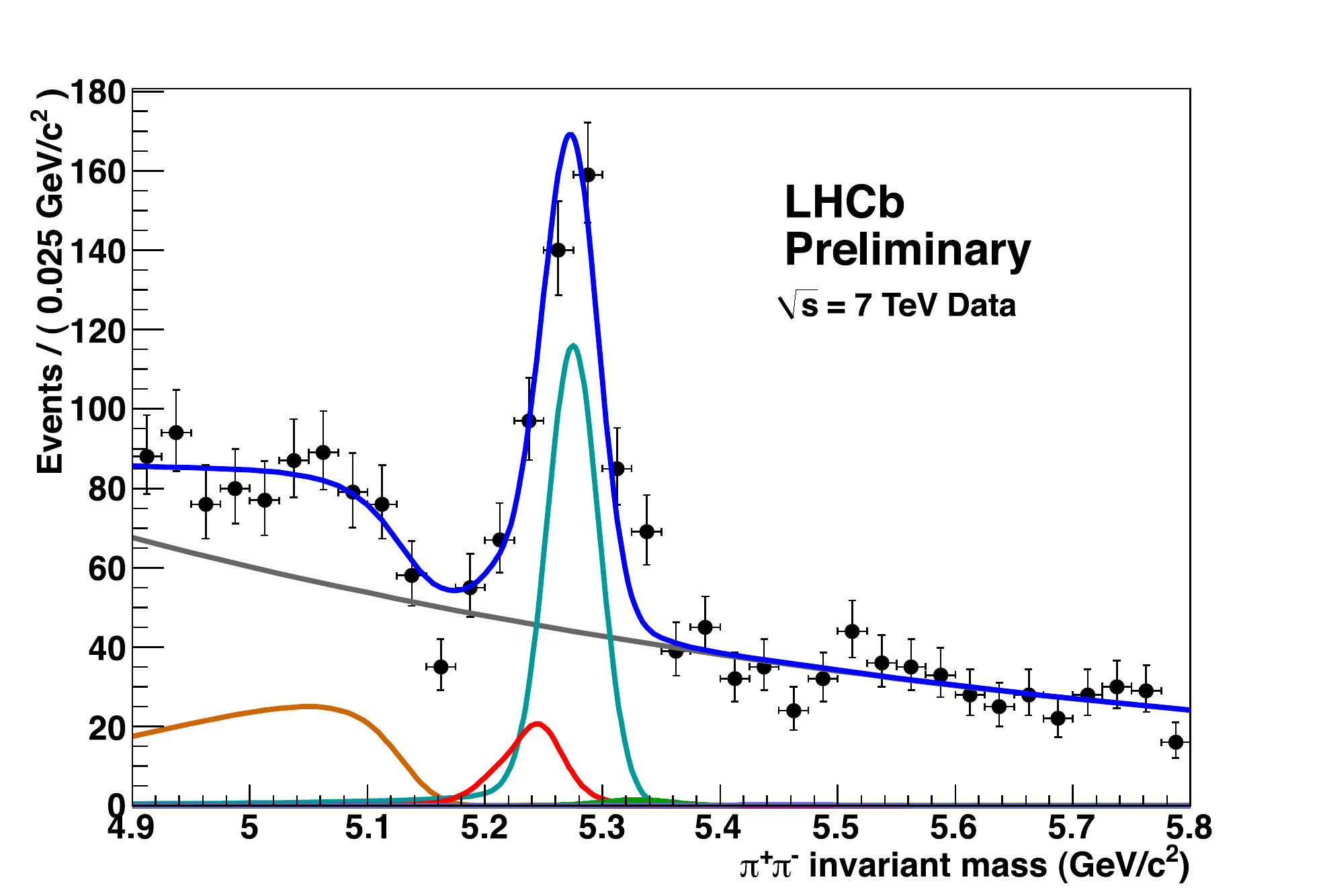}
\includegraphics[width=68mm,height=55mm]{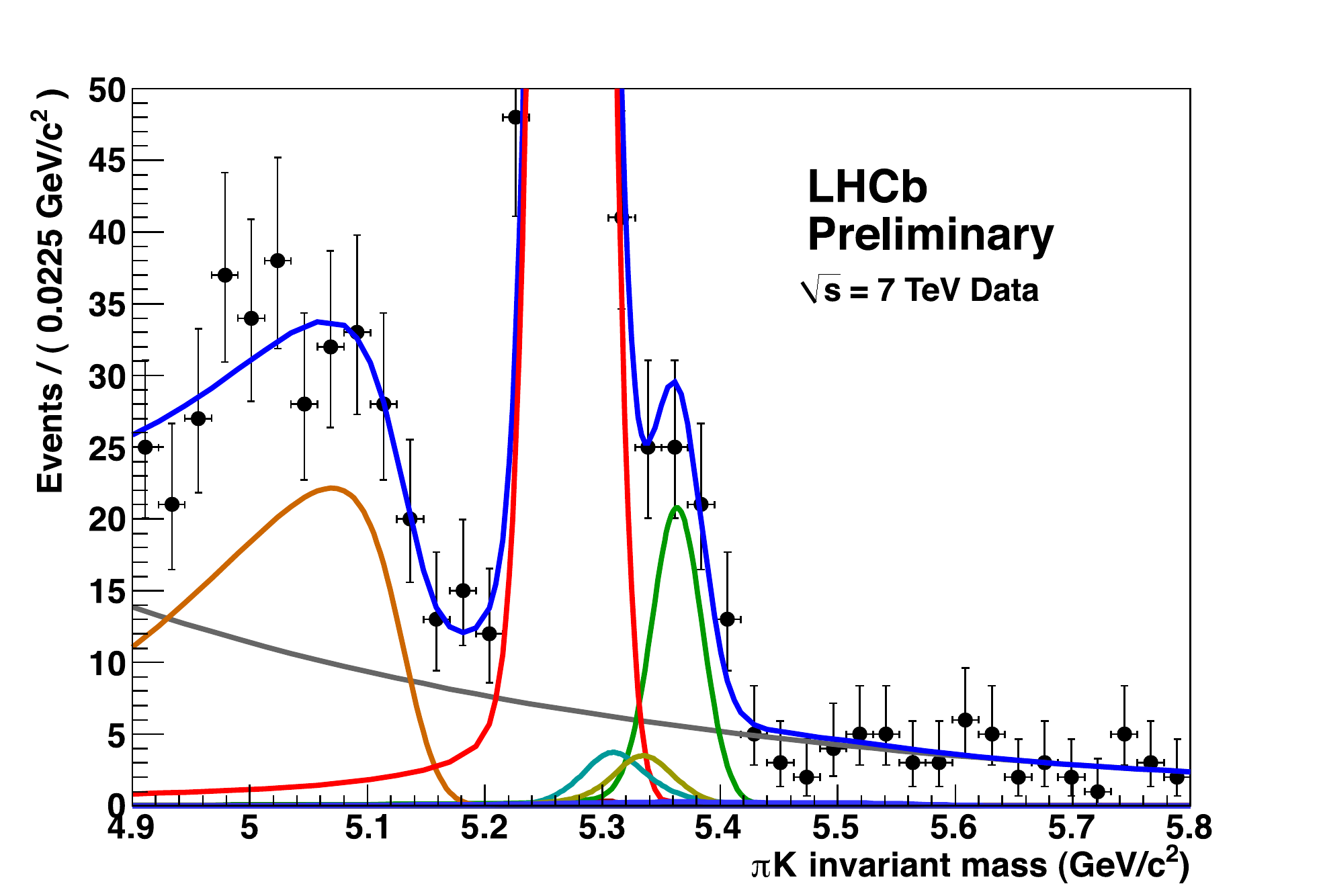}
\hfil
\includegraphics[width=68mm,height=55mm]{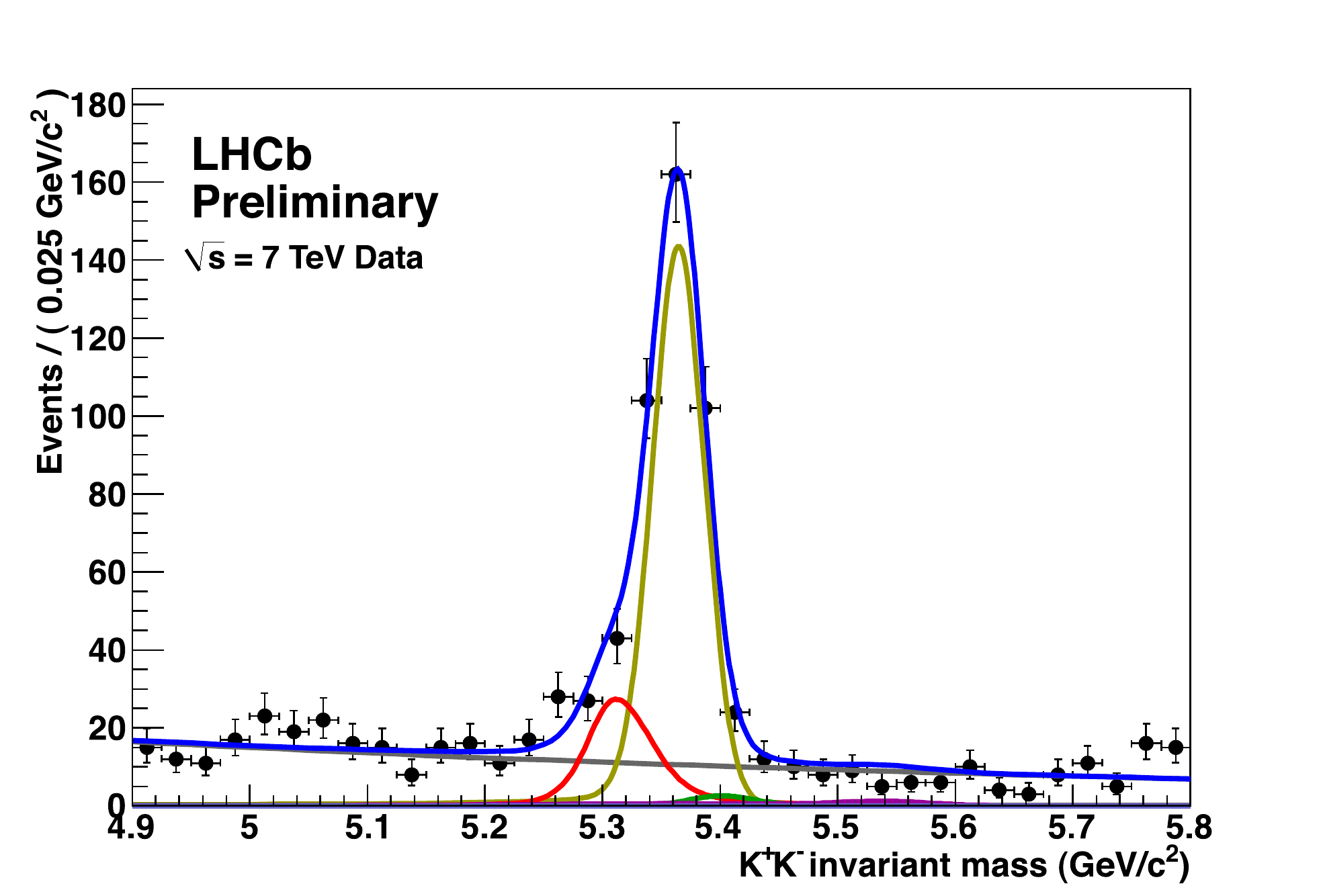}
\caption{Invariant mass spectra for $B^0 \rightarrow K^+\pi^-$ (top, left), $B^0 \rightarrow \pi^+\pi^-$ (top, right), $B^0_s \rightarrow \pi^+K^-$ (bottom, left, green line), and $B^0_s \rightarrow K^+K^-$ (bottom, right). The analysis is based on $\sim37~\rm{pb}^{-1}$ collected during the 2010 run.\label{btohh}}
\end{figure}

The RICH system provides good particle identification over the entire momentum range. The efficiency for kaon identification is measured to be larger than 90\%, with a corresponding pion misidentification rate $\lesssim$10\% for momenta below $70~\gevc$, which is the relevant range for most hadrons from $b$ decays. As an illustration of the RICH capabilities to disentangle the various $B$ decay modes, Fig.~\ref{btohh} shows the mass distributions obtained for a set of charmless charged two-body $B$ hadron decay modes, namely  $B^0 \rightarrow K^+\pi^-$, $B^0 \rightarrow \pi^+\pi^-$,  $B^0_s \rightarrow \pi^+K^-$, and $B^0_s \rightarrow K^+K^-$\cite{ref:b2hh}. 
\subsection{Running experience}
During the 2010 run, LHCb collected an integrated luminosity of $\sim37~\rm{pb}^{-1}$  at $\sqrt{s}=7$~TeV with an overall efficiency of $\sim$90\%. The performance of the LHC  improved progressively during the course of the year and most of the integrated luminosity was collected with the last few fills. Initially, the LHC operated with a reduced number of colliding bunches ($<50$), which allowed LHCb to run with a very loose trigger. However, towards the end of the run, the LHC reached an impressive peak luminosity of $\simeq 1.6\times10^{32}\cms$ by colliding a number of bunches eight times smaller than nominal (344 instead of 2622) but achieving outstanding beam characteristics, $\it{i.e.}$, $\sim1.1\cdot10^{11}$ protons-per-bunch, a squeezing function $\beta$*~=~3.5~m and a normalized emittance $\epsilon_{\rm{N}}=2.4~\mu$m. This, however, implied an average number of visible proton-proton interactions per bunch crossing $\mu\gtrsim2.5$, a factor of six above the LHCb design value, and running conditions extremely challenging for the trigger, the offline reconstruction and the data processing. LHCb was able to cope with these conditions and data were taken at the highest luminosity available from LHC at all times. This required increasing the CPU capacity of the event-filter farm, which had turned-out to be the main limitation,  and introducing cuts on the  hit multiplicities of the sub-detectors used by the pattern-recognition algorithms to reject events with a large number of pile-up interactions, while loosing a minimal amount of luminosity. Half of the bandwidth ($\sim$1kHz) was attributed to the muon trigger lines, which rely on very low $\PT$ cuts,  to be highly efficient for the \BSMUMU analysis. 

The very good experience from the 2010 run, in which LHCb was successfully operated at an instantaneous luminosity much higher than anticipated, opens interesting prospects for the 2011-2012 run. A procedure of 'Luminosity leveling'  based on displacing the beams in the vertical direction is currently being defined with the LHC. This will imply less pile-up at the start of the fill and a more constant behaviour in time, allowing LHCb to collect the maximum integrated luminosity under optimal conditions.

\section{Selected physics results and prospects}
The data collected in 2010 have already allowed LHCb to realize a large number of significant measurements. Here only a selection of them will be reviewed.

The study of bound \cc\ and \bb\ states decaying into di-muons is a natural early physics topic for LHCb.  Due to its forward acceptance and dedicated trigger, LHCb has the largest \Jpsi\ sample at the LHC. This was used to study \Jpsi\ hadroproduction, which is not so well understood theoretically, despite the considerable progress made in recent years~\cite{qwgdoc}. 

\begin{figure} [tbh]
\includegraphics[width=68mm,height=55mm]{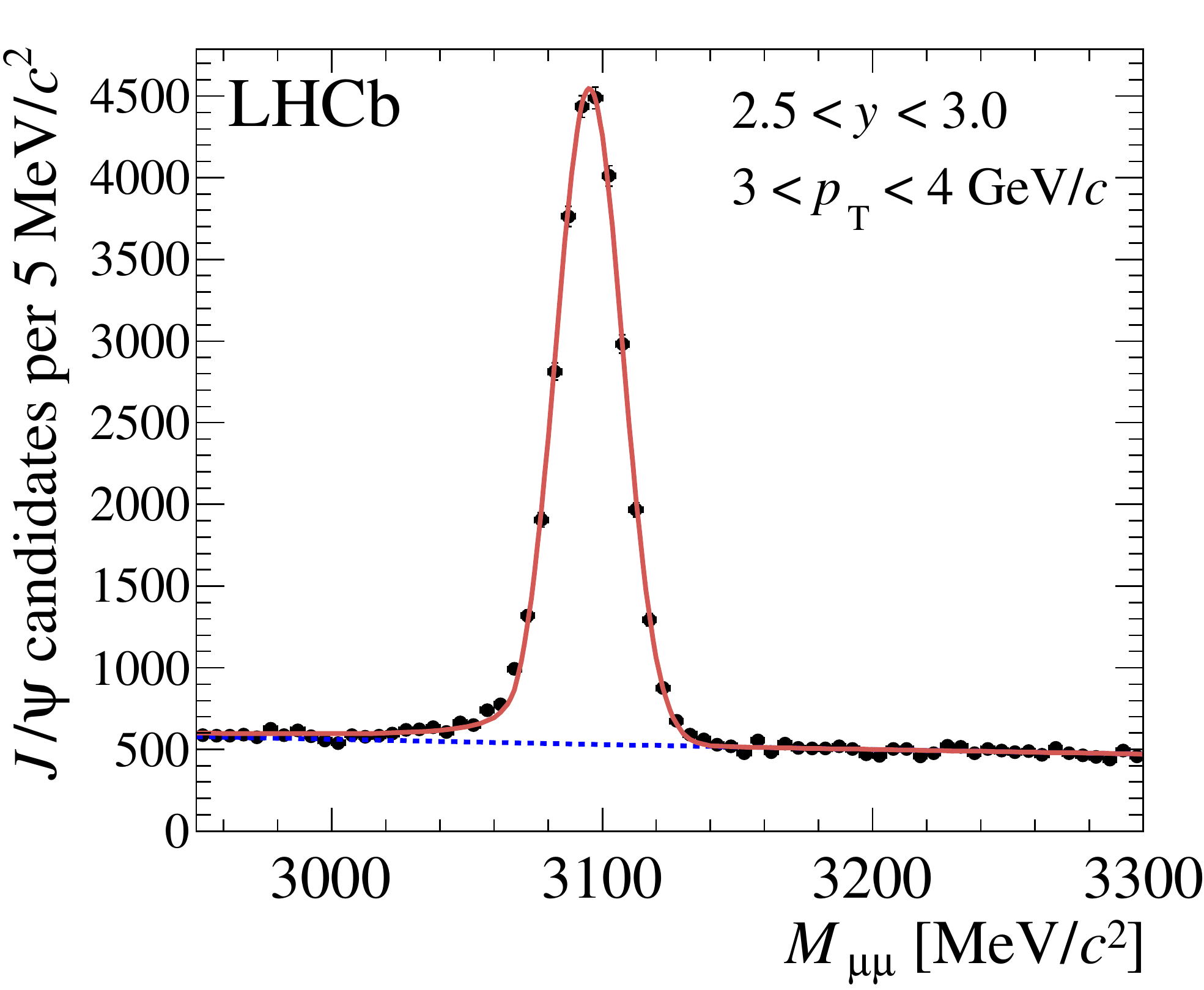}
\hfil
\includegraphics[width=68mm,height=53mm]{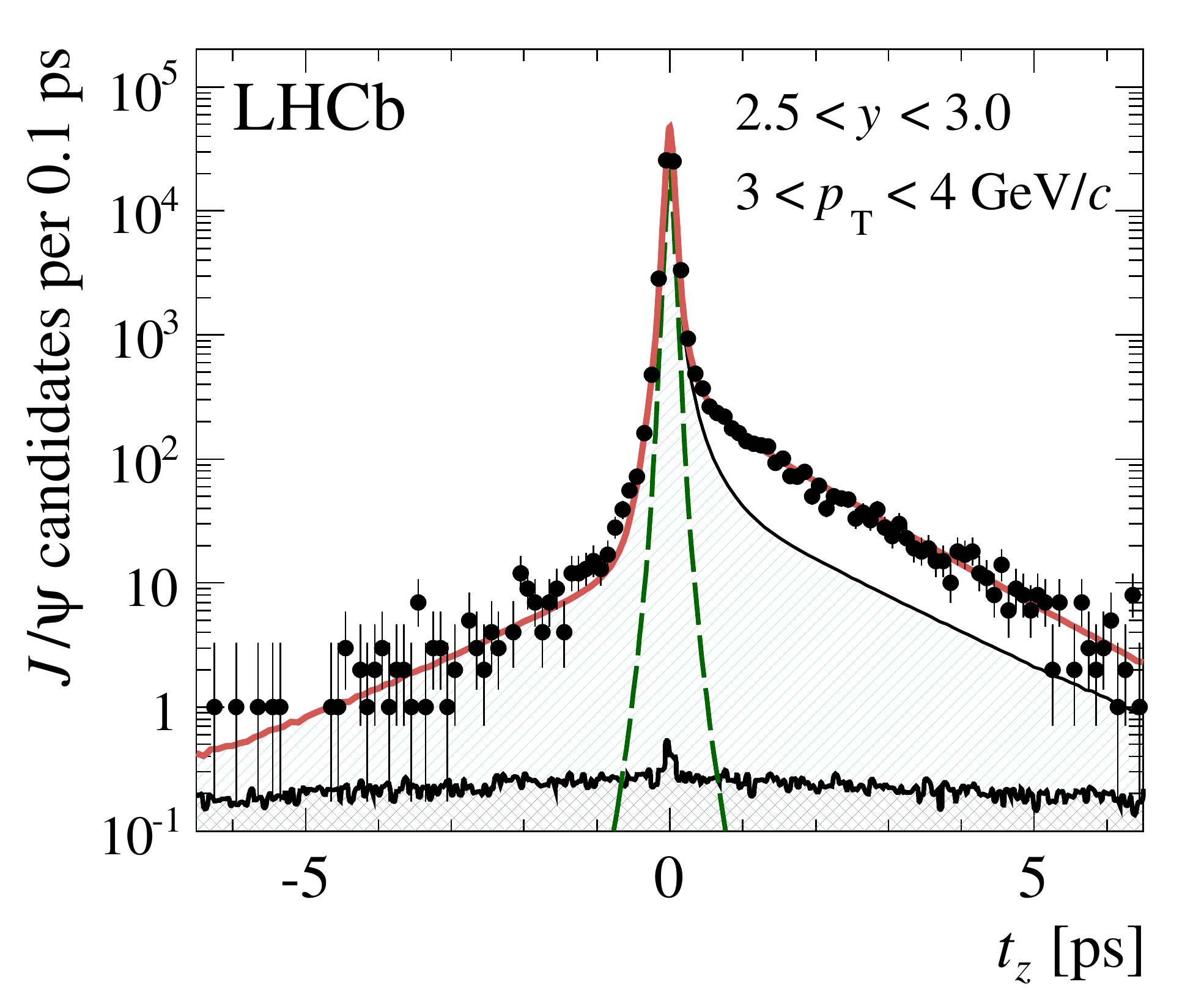}
\caption{ Di-muon mass distribution (left) and $t_z$ distribution (right), with fit results superimposed, for a specific  bin ($3 < \PT <4\,\gevc$,\,$2.5<y<3.0$). On the mass distribution, the solid red line is the total fit function, where the signal is described by a Crystal Ball function, and the dashed blue line represents the exponential background function.  The fit gives a mass resolution of $12.3\pm0.1~\mevcc$. On the $t_z$ distribution, the solid red line is the total fit function, the green dashed line is the prompt contribution, and the single-hatched area is the background component.} \label{fig:tzresult}
\end{figure}

The analysis is based on $5.2~\rm{pb}^{-1}$ of data in which the \Jpsi\ mesons are reconstructed in the decay mode $\Jpsi\rightarrow \mu^+ \mu^-$~\cite{ref:jpsi}. \Jpsi\ mesons from $b$-hadron decays, which tend to be produced away from the primary vertex,  are separated  from the prompt \Jpsi\ component by exploiting a variable defined as $t_z = \frac{(z_{\Jpsi}-z_{\rm PV}) \times M_{\Jpsi}}{p_z}$, where $z_{\Jpsi}$ and $z_{\rm PV}$ are the positions along the $z$-axis (defined along the beam axis) of the \Jpsi\ decay vertex and of the primary vertex; $p_z$ is the measured \Jpsi\ momentum in the $z$ direction and $M_{\Jpsi}$ the nominal \Jpsi\ mass.  The $t_z$ variable provides  a good estimate of  the $b$-hadron decay proper time. The differential production cross-section of both prompt \Jpsi\ and \fromb\ is measured as a function of the \Jpsi\ transverse momentum $\PT$ and rapidity $y$ in the fiducial region  $\PT\in[0\,;14]\gevc$  and  $y\in[2.0\,;4.5]$ in bins of $\Delta y= 0.5$ and $\Delta \PT= 1~\gevc$.  As an example, Fig.~\ref{fig:tzresult} shows for one specific bin ($3 < \PT <4\,\gevc$, \,$2.5<y<3.0$) the di-muon mass distribution (left)  and the $t_z$ distribution (right), in which one can clearly distinguish the prompt component at $t_z$=0 and  the \fromb\ component characterised by an exponential decay. The integrated cross-section for prompt \Jpsi\ production in the defined fiducial region, summing over all bins of the analysis, is $\sigma_{\rm prompt\,\Jpsi}= \, 10.52\pm 0.04\pm 1.40^{+1.64}_{-2.20}\,\microb$. The result is quoted assuming unpolarised \Jpsi\ and the last error indicates the uncertainty related to this assumption.The integrated cross-section for the production of \fromb\ in the same fiducial region is $\sigma_{\fromb} =  \, 1.14 \pm 0.01\pm 0.16\,\microb$. By extrapolating this last measurement to the full polar angle using the LHCb Monte Carlo simulation based on  {\sc Pythia} 6.4~\cite{pythia} and the average branching fraction of inclusive $b$-hadron decays to \Jpsi\ measured at LEP\cite{delphibtojpsi} ${\cal B}(b\to\Jpsi X)=(1.16\pm0.10)\%$, one obtains that the  \bb\ cross-section in $pp$ collisions at $\sqrt{s}=7$~TeV  is $\sigma(pp \to b\overline{b} X) = 288\pm 4 \pm 48 \, \microb$. These cross-section results are systematics limited, with the largest uncertainty  (10\%) being due to the absolute luminosity scale, dominated by the uncertainty in the knowledge of the LHC proton beam currents. However, this uncertainty is now reduced to $\sim3.5\%$\cite{ref:lumi}. The \Jpsi\ analysis will be repeated on a larger data sample to extract a measurement of the \Jpsi\ polarisation and benefit from the reduced systematic uncertainties.

Also in the domain of bound \cc\ and \bb\ states decaying into di-muons, LHCb made the first observation of double \Jpsi\ production at the LHC~\cite{ref:doubleJpsi} and released preliminary results on $\Upsilon$ production~\cite{ref:upsilon}. The $\Jpsi\rightarrow \mu^+ \mu^-$ signature was also used to isolate a sample of $B_c$ candidates~\cite{ref:Bc}, to study  $\psi(2S)$ production by exploiting the two decay modes $\psi(2S)\rightarrow\mu^+\mu^-, ~\Jpsi(\mu^+\mu^-)\pi^+\pi^-$ and to measure the mass of the $X(3872)$ state~\cite{ref:X}.

\begin{figure} [tbh]
\includegraphics[width=15.0 cm]{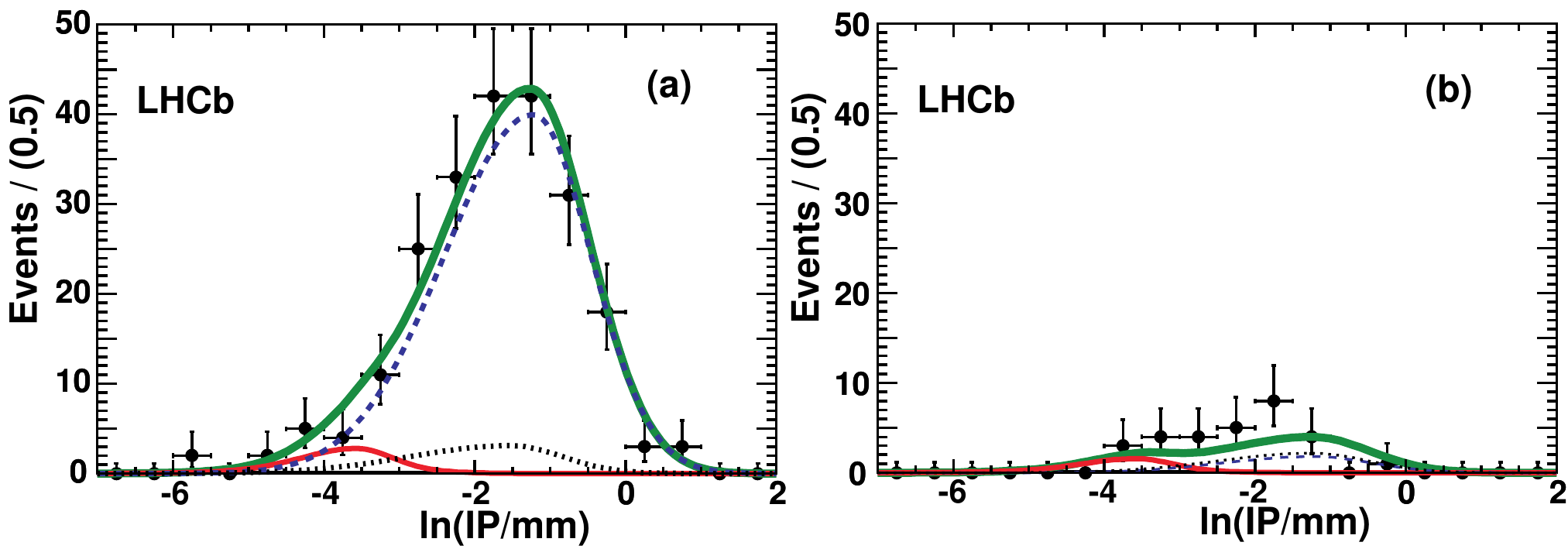}
\caption{ Natural logarithm of the $D^0$ impact parameter for (a) right-sign and (b) wrong-sign $D^0$-muon candidate combinations. The dotted curves show the $D^0$ sideband backgrounds, the dashed curve the signal and the thick solid curves the totals.} \label{fig:semilept} 
\end{figure}

The first LHCb measurement of the \bb\ cross-section~\cite{ref:semilept} was in fact performed by analyzing $b$ hadrons decays into final states containing a $D^0$ meson and a muon. A clean sample of $D^0\rightarrow K\pi$ was isolated and the impact parameter of the $D^0$ with respect  to the primary vertex used to separate prompt $D^0$ mesons from those originating from $b$-flavoured hadron decays. Signal events are characterised  by the sign of the charge of the muon being the same as the charge of the kaon in the $D^0$ decay, while wrong-sign combinations are highly suppressed. As an illustration, Fig.~\ref{fig:semilept} shows the natural logarithm of the $D^0$ impact parameter  for (a) right-sign and (b) wrong-sign $D^0$-muon candidate combinations.
Based on a sample of $15~\rm{nb}^{-1}$, the cross-section to produce a $b$-flavoured hadron $H_b$ (averaging over $b$-flavoured and $\bar b$-flavoured hadrons) is  $\sigma(pp\rightarrow H_b X) = 75.3 \pm 5.4 \pm 13.0\microb$ in the pseudorapidity interval $2 < \eta < 6$ over the entire range of $\PT$, assuming the $b$ fragmentation fractions measured at LEP. By extrapolating to the full $\eta$  region, one measures $\sigma(pp \to b\overline{b} X) = 284\pm 20 \pm 49 \, \microb$, in remarkable agreement with the result obtained with the \Jpsi\ sample.

The decay \BSMUMU has been identified as a very interesting potential constraint on the parameter space of models for physics beyond the SM. The BR for this decay is computed to be very small in the SM: BR(\BSMUMU)$=(3.2\pm0.2)\times{10^{-9}}$\cite{ref:bsmumuth}, but could be enhanced in certain NP scenarios. For example, in the MSSM, this branching ratio is known to increase as the sixth power of $\rm{tan}\beta=\nu_{\rm u}/\nu_{\rm d}$, the ratio of the two vacuum expectation values. Any improvement to this limit is therefore particularly important for models with large $\rm{tan}\beta$. The best published 95\%~CL upper limits to the \BSMUMU branching ratio are from CDF:  BR(\BSMUMU)$<5.8\times 10^{-8}$\cite{ref:CDFbsmumu} and D0: BR(\BSMUMU)$<5.1\times 10^{-8}$\cite{ref:D0bsmumu}, based on integrated luminosities of  $2~\rm{fb}^{-1}$ and $6.1~\rm{fb}^{-1}$, respectively. Preliminary results from CDF with $3.7~\rm{fb}^{-1}$~\cite{ref:CDFbsmumu_prel}   lower the limit  to  $4.3\times 10^{-8}$. 

\begin{figure} [tbh]
\includegraphics[width=68mm,height=55mm]{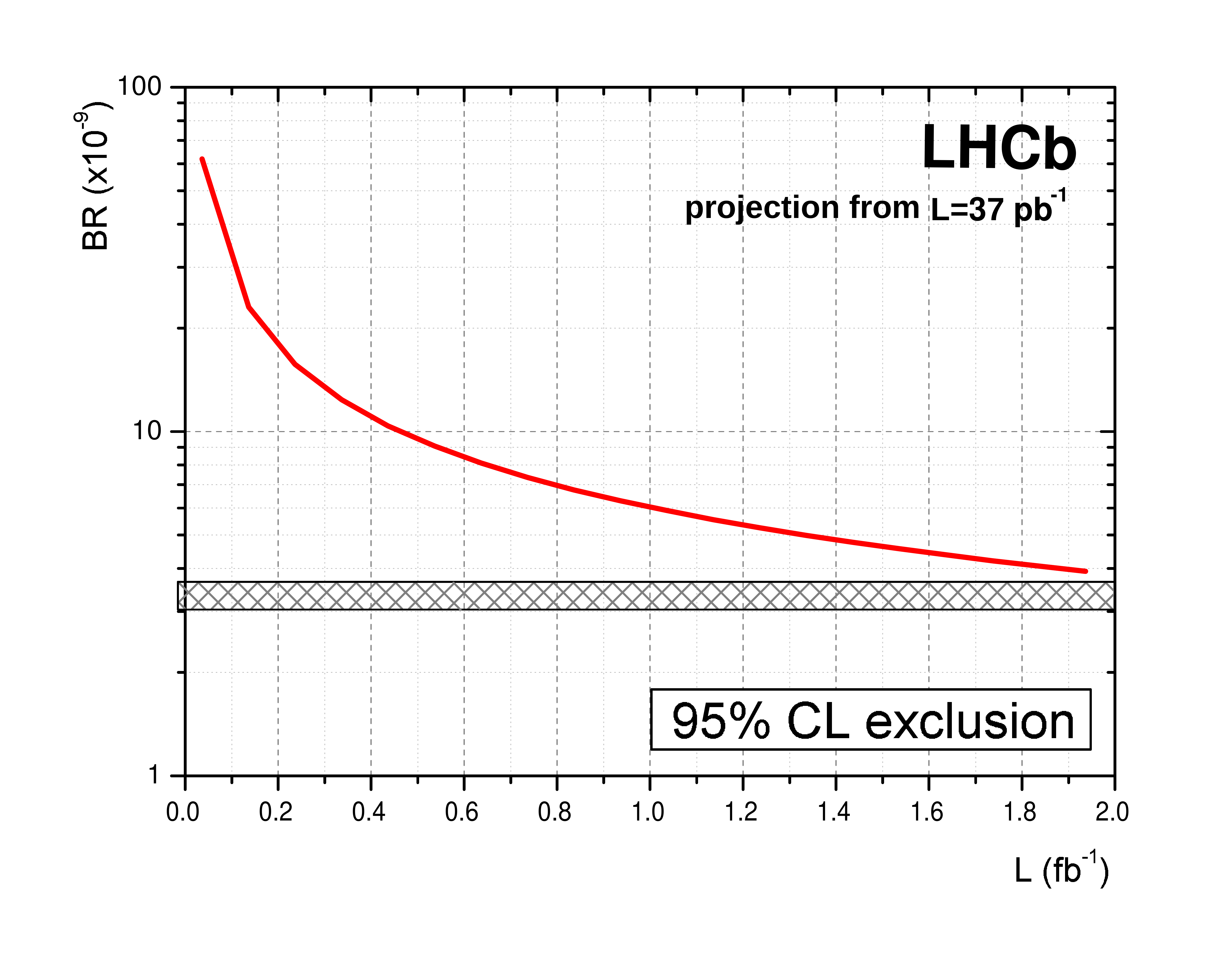}
\hfil
\includegraphics[width=68mm,height=56mm]{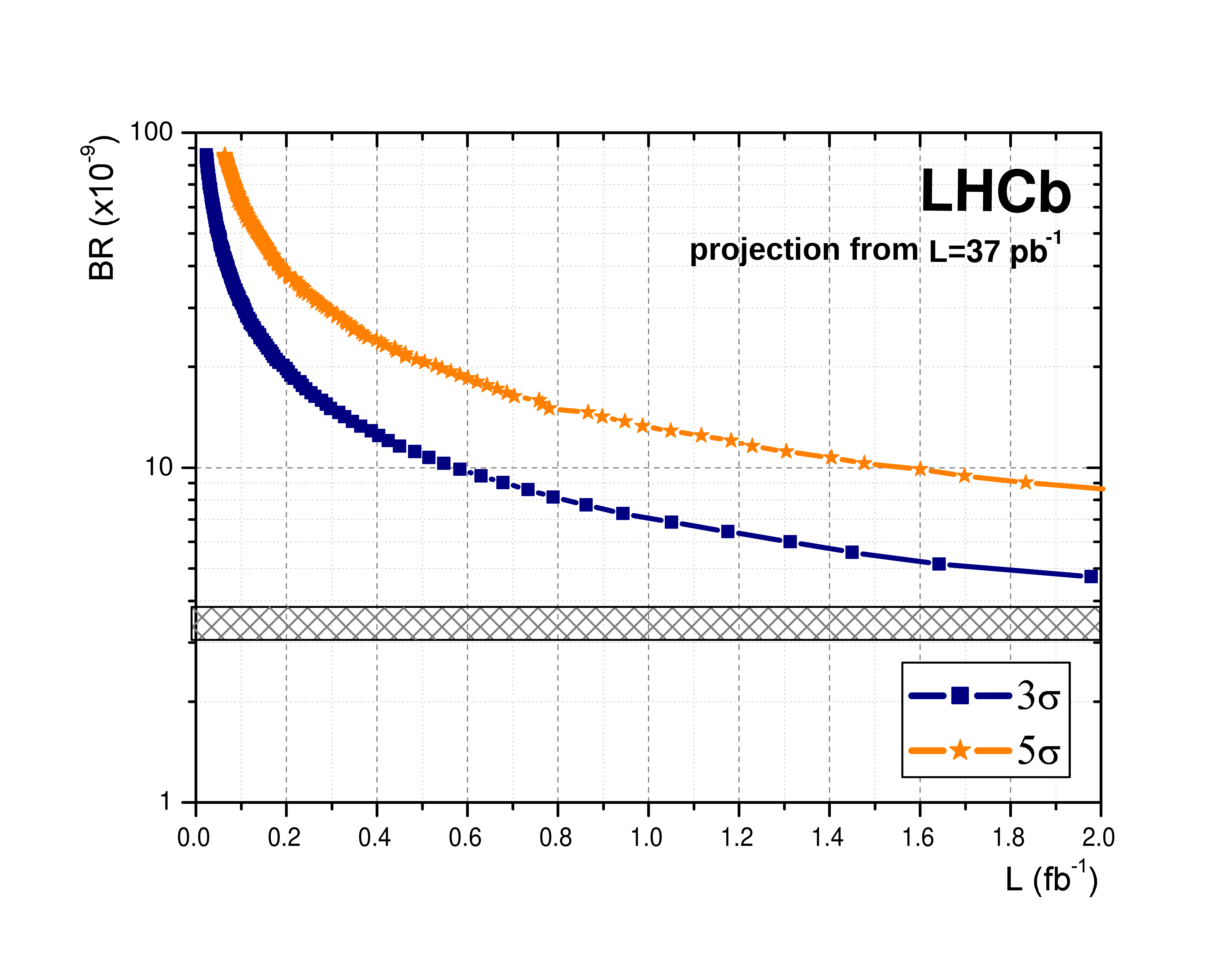}
\caption{ Expected  95\% CL upper limit of BR(\BSMUMU) (left) in the absence of signal and  BR(\BSMUMU) at which a 3 or 5~$\sigma$ observation  is expected (right) as a function of integrated luminosity. The grey band is the SM value. } \label{fig:bsmumu} 
\end{figure}

LHCb has recently reported results for the search for \BSMUMU (and \B0MUMU, expected to be thirty times smaller, for which the analysis is the same) based on the full 2010 data sample of $37~\rm{pb}^{-1}$\cite{ref:LHCbbsmumu}. The analysis is based on the use of control channels to minimize the dependence on MC simulation.  The main issue is the rejection of the background, largely dominated  by random combinations of two muons originating from two distinct $b$ decays. This background can be kept under control by exploiting the excellent LHCb vertexing capabilities, and mass resolution ($\sigma=26.7 \pm 0.9\mevcc$). Three normalisation channels are used to calculate the BR, $\it i.e.$,   \BJMUK, \BSJMUPHIK and \BKPI, which allows us to cancel out many systematic uncertainties in the ratio of the efficiencies and to eliminate any dependence on absolute luminosity and \bb\ cross-section. The event selection for these channels is specifically designed to be as close as possible to the signal selection. Each selected  \BMUMU candidate is assigned a probability to be signal or background in a two-dimensional probability space defined by the di-muon invariant mass and a geometrical likelihood (GL). The GL is based on several geometrical properties of the decay, such as lifetime, transverse momentum and impact parameter of the $B$ candidate, muon isolation, etc. The di-muon invariant mass and GL probability density functions for both signal and background are determined from data. %Figure~\ref{fig:bsmumu}  shows the observed distribution of $\rm CL_s$ as a function of the assumed branching ratio for \BSMUMU (a) and \B0MUMU (b). 
The observed upper limits are: BR(\BSMUMU)$<4.3 (5.6) \times 10^{-8}$ at 90\% (95\%) CL and BR(\B0MUMU)$<1.2 (1.5) \times 10^{-8}$ at 90\% (95\%) CL. So, thanks to the large acceptance and trigger efficiency, the LHCb sensitivities are already similar to the existing limits from CDF and D0.
Figure~\ref{fig:bsmumu} illustrates the expected  95\%~CL upper limit of BR(\BSMUMU) (left) in the absence of signal and observation at 3 or 5~$\sigma$ (right) as a function of integrated luminosity, based on the performance achieved with the 2010 data. This shows that with the data expected during the 2011-2012 run, LHCb will obtain a 90\% exclusion limit approaching the SM and explore branching ratios in the range $\sim$4-10~$10^{-9}$.

\begin{figure} [tbh]
\begin{center}
\includegraphics[width=10.0 cm]{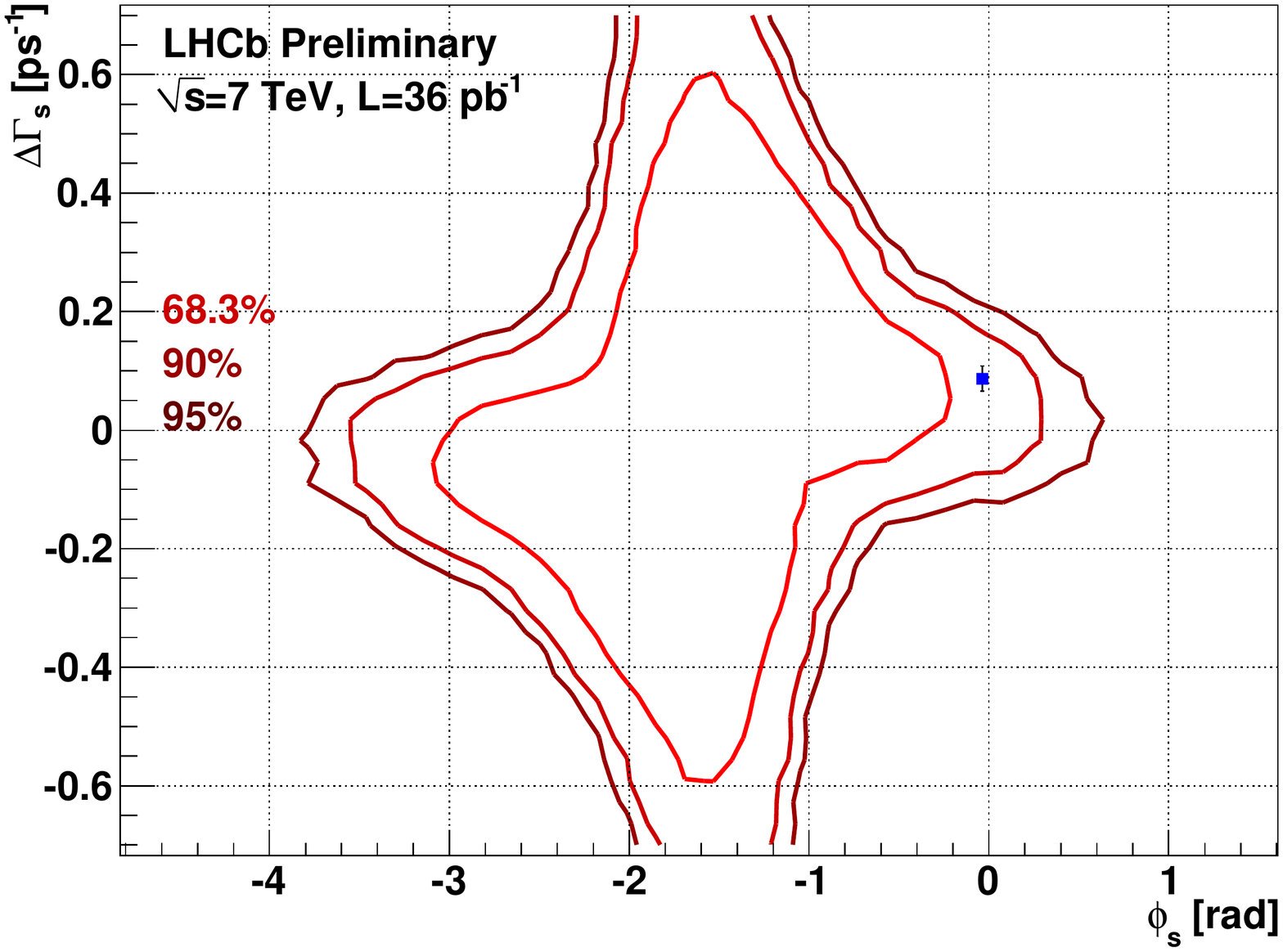}
\end{center}
\caption{ Feldman-Cousins confidence regions in the  ($\phi_s, \Delta\Gamma_s$) plane. The blue square close to $\phi_s=0$ give the SM prediction.} \label{fig:phi_s} 
\end{figure}

A flavour tagged, angular analysis of the decay \BJPSIPHI allows the determination of the CP-violating phase $\phi_s$. In the SM this phase is predicted to be  $\simeq-2\beta_s$, where $\beta_{\rm s}$ is the smaller angle of the "$b$-$s$ unitarity triangle". The indirect determination via fits to the experimental data gives $2\beta_s= (0.0363 \pm 0.0017)$~rad\cite{ref:CKM}. However, NP could significantly modify this prediction, if new particles contribute to the \BBar box diagram. In fact, the CDF and D0 collaborations\cite{ref:CDFphis, ref:D0phis} have reported measurements of the \BS mixing phase using approximately 6500 \BJPSIPHI candidates from $5.2~\rm{fb}^{-1}$ and 3400 \BJPSIPHI candidates  from $6.1~\rm{fb}^{-1}$, respectively. Both results are compatible with the SM expectation at slightly more than one standard deviation in the ($\phi_s, \Delta\Gamma_s$) plane. However, the measurements still suffer from large statistical uncertainties. LHCb has recently demonstrated~\cite{ref:LHCbphis} to have the capability to significantly improve the existing experimental knowledge of this phase thanks to the large signal yield (836 events for $36~\rm{fb}^{-1}$), and the excellent proper time resolution to resolve fast \BS oscillations ($\sim$50~fs). The analysis uses an opposite-side  flavour tagger based on four different signatures, namely high $\PT$ muons, electrons and kaons, and the net charge of an inclusively reconstructed secondary vertex, with an effective tagging power of $\sim2.2\%$. The results obtained on the 2010 data sample are presented in Fig.~\ref{fig:phi_s} as two-dimensional confidence level regions in the ($\phi_s, \Delta\Gamma_s$) plane, following the prescription of Feldman-Cousins~\cite{ref:Feldman}. Projected in one dimension, we find $\phi_s\in[-2.7\,;-0.5]$~rad at 68\% CL.
Based on the sensitivity demonstrated with the 2010 data, LHCb will produce the world best $\phi_s$ measurement with the $\sim1~\rm{fb}^{-1}$ expected by the end of  2011. Moreover, adding a same-side kaon tagger should significantly improve the sensitivity.

There are several other measurements that I have not been able to cover in this review, such as the ratio of the fragmentation functions $f_s/f_d$~\cite{ref:bfrag}, (where $f_s$ and $f_d$ describe the probability that a $b$ quark will fragment into a $B_{s,d}$ meson), which is crucial for a precise determination of any $B^0_s$ branching ratio at the LHC, including BR(\BSMUMU), or in areas beyond heavy flavour, such as the production of $W$ and $Z$ in the forward direction~\cite{ref:WZ}.  It is clear that the experiment is performing according to expectations. It has made several new observations and in several key analyses has already obtained sensitivities very similar to those of the $B$-factories and Tevatron. The prospects to be able to probe NP during the 2011-2012 run are therefore very exciting.

\section{Conclusions}
The large \bb\ production cross section at the LHC provides a unique opportunity to study in detail CP violation  $b$ decays with the LHCb detector. In particular, production of \BS mesons could play a crucial role in disentangling effects originating from NP.  LHCb has already performed many high-quality measurements with the data collected in 2010. An integrated luminosity of $\sim1~\rm{fb}^{-1}$ , as expected in 2011, will allow LHCb to realize a number of very significant $b$-physics measurements, with the potential of revealing NP effects.
\begin{acknowledgement}
I would like to thank the Organizers of the Corfu Summer Institute for their kind invitation, my LHCb colleagues for providing the material discussed here and, in particular, Guy Wilkinson and Patrick Robbe for their careful reading of this article.
\end{acknowledgement}


\begin{thebibliography}{[1]}
\bibitem{ref:reopt} The LHCb Collaboration, ``LHCb Reoptimized Detector, Design and Performance",  CERN/LHCC-2003-030 LHCb TDR~9~(2003)
\bibitem{ref:det} The LHCb Collaboration, ``The LHCb Detector at the LHC", Journal of Instrumentation, JINST 3 S08005 (2008)
\bibitem{ref:b2hh} The LHCb Collaboration, ``Measurement of direct CP violation in charmless charged two-body $B$ decays at LHCb", LHCb-CONF-2011-011 (2011) 
\bibitem{qwgdoc} N.~Brambilla {\it et al.}, ``Heavy quarkonium: progress, puzzles, and opportunities'',  Eur. Phys. J. C {\bf71} (2011) 1534, \href{http://arxiv.org/abs/1010.5827}{arXiv:1010.5827 [hep-ph]}.
\bibitem{ref:jpsi} The LHCb Collaboration, ``Measurement of \Jpsi\ production in {\it pp} collisions at  $\sqrt{s}=7$~TeV", Eur. Phys. J. C {\bf71} (2011) 1645, \href{http://arxiv.org/abs/1103.0423}{arXiv.org/abs/1103.0423}
\bibitem{pythia} T.~Sj\"{o}strand,  S.~Mrenna and P.~Z.~Skands, ``{\sc Pythia} 6.4 physics and manual'', version 6.422, J. High Energy Phys. {\bf 0605} (2006) 026, \href{http://arxiv.org/abs/hep-ph/0603175}{arXiv:1103.0423}
\bibitem{delphibtojpsi} The DELPHI Collaboration, P.~Abreu {\it et al.}, ``\Jpsi\ production in the hadronic decays of the $Z$'', Phys. Lett. B {\bf 341} (1994) 109;
The L3 Collaboration, O.~Adriani {\it et al.}, ``$\chi_c$ production in hadronic $Z$ decays'', Phys. Lett. B {\bf 317} (1993) 467;
The ALEPH Collaboration, D.~Buskulic {\it et al.}, ``Measurements of mean lifetime and branching fractions of $b$ hadrons decaying to \Jpsi'', Phys. Lett.  B {\bf 295} (1992) 396
\bibitem{ref:lumi} J.Panman, ``Recent results from LHCb", Proceedings for 23rd International Workshop on Weak Interactions and Neutrinos,\href{http://cdsweb.cern.ch/record/1350263}{LHCb-PROC-2011-030}
\bibitem{ref:doubleJpsi}  The LHCb Collaboration, ``Observation of double \Jpsi\ production in proton-proton collisions at a centre-of-mass energy of $\sqrt{s}$=7~TeV", LHCb-CONF-2011-009
\bibitem{ref:upsilon}  The LHCb Collaboration, ``Measurement of the $\Upsilon(1S)$ production cross-section at  $\sqrt{s}$=7~TeV in LHCb", LHCb-CONF-2011-016
\bibitem{ref:Bc} The LHCb Collaboration, ``Measurement  of the $B^+_c$ to $B^+$ production cross-section ratio at  $\sqrt{s}$=7~TeV in LHCb", LHCb-CONF-2011-017
\bibitem{ref:X}  The LHCb Collaboration, ``Measurement of the $X(3872)$ mass with first LHCb data", LHCb-CONF-2011-021
\bibitem{ref:semilept} The LHCb Collaboration,  ``Measurement of $\sigma(pp\to b\overline{b}X)$ at $\sqrt{s}$=7~TeV in the forward region'',  Phys. Lett. B {\bf 694} (2010) 209, \href{http://arxiv.org/abs/1009.2731}{arXiv:1009.2731 [hep-ex]}.
\bibitem{ref:bsmumuth} A.~Buras,  ``Minimal flavour violation and beyond: Towards a flavour code for short distance dynamics", \href{http://arxiv.org/abs/1012.1447 }{arXiv:1012.1447} (2010)
\bibitem{ref:CDFbsmumu} The CDF Collaboration, ``Search for \BSMUMU and \B0MUMU decays with $2~\rm{fb^{-1}}$ of $p\bar p$ CollisionsÓ, Phys. Rev. Lett. {\bf 100} (2008) 101802
\bibitem{ref:D0bsmumu} The CDF Collaboration, ``Search for the rare decay \BSMUMUÓ, Phys. Lett. B {\bf 693} (2010) 539
\bibitem{ref:CDFbsmumu_prel} The CDF Collaboration, CDF Public Note 9892, (2009) 
\bibitem{ref:LHCbbsmumu} The LHCb Collaboration  ``Search for  the rare decays \BSMUMU and \B0MUMU", \href{http://arxiv.org/pdf/1103.2465v2 }{arXiv:1103.2465}, Phys. Lett.  B {\bf 699} (2011) 330
\bibitem{ref:CKM} J.~Charles {\it et al.}, (CKMfitter group), Eur. Phys. J. C{\bf 41}, 1-131 (2005), hep-ph/0406184, updated results and plots available at \href{http://ckmfitter.in2p3.fr/}{http://ckmfitter.in2p3.fr} 
\bibitem{ref:CDFphis} The CDF Collaboration, public note CDF/ANAL/BOTTOM/PUBLIC/10206 (2010)
\bibitem{ref:D0phis} The D0 Collaboration, D0 Conference note 6098-CONF (2010)
\bibitem{ref:LHCbphis} The LHCb Collaboration, ``Tagged time-dependent angular analysis of \BJPSIPHI decays with the 2010 LHCb data", LHCb-CONF-2011-006 
\bibitem{ref:Feldman} G.J.~Feldman and R.D. Cousins, ``A Unified Approach to the Classical Statistical Analysis of Small Signals" , Phys. Rev. D {\bf 57} (1998) 3873 
\bibitem{ref:bfrag} The LHCb Collaboration, ``Measurement of the relative yields of the decay modes $B^0 \to D^- \pi^+$ , $B^0 \to D^- K^+$, $B^0_s \to D^-_s \pi^+$, and determination of $f_s/f_d$ for 7 TeV $pp$ collisions", LHCb-CONF-2011-013
\bibitem{ref:WZ} The LHCb Collaboration, ``$W$ and $Z$ production at $\sqrt{s}$=7~TeV  with the LHCb experiment", LHCb-CONF-2011-012
\end{thebibliography}
\end{document}